\documentclass[aps, pra, 10pt, twocolumn, superscriptaddress, nofootinbib]{revtex4-2}
\usepackage[english]{babel}
\usepackage{graphicx}
\usepackage{dcolumn}
\usepackage{bm}
\usepackage{lipsum}
\usepackage{amsmath}
\usepackage{gensymb}
\usepackage{amsmath}
\usepackage{amssymb}
\usepackage{physics}
\usepackage{textcomp}

\begin{document}

\preprint{APS/123-QED}

\title{Inverse design of polarization-analyzing gratings}

\author{Kexin Bian}
\author{Lisa W. Li}
\author{Alex R. Fiani}
\author{Noah A. Rubin}\email{noahrubin@ucsd.edu}
\affiliation{Department of Electrical and Computer Engineering, University of California San Diego, La Jolla, CA 92093 USA}

\date{\today}

\begin{abstract}
Metasurface polarization analyzing gratings have attracted increasing interest for polarization management and sensing in optical systems. Traditional strategies for the design of these gratings are based on a formalism combining the Jones calculus with Fourier optics.  While this approach has achieved notable success in numerous applications, its validity is restricted to the paraxial regime. Here we introduce a differentiable topology optimization framework for these gratings which is free of this constraint. We show that this framework can yield designs which outperform the traditional approach, especially at high angles. Additionally, we provide preliminary experimental validation to corroborate these results.
\end{abstract}

\maketitle

\section{Introduction} 
\vspace*{-3pt}

Polarization describes the direction of oscillation of the electromagnetic field of light. As a fundamental property of electromagnetic waves, polarization and its measurement play a critical role in a wide range of applications, including optical communication, imaging, and wavefront sensing. Conventional methods for polarization analysis usually rely on sequential measurements using optical components such as waveplates and polarizers, which has limited the compactness and acquisition speed of polarization imaging systems~\cite{Tyo2006}.

Traditionally, acquiring all Stokes parameters in a snapshot measurement with a compact optical setup has presented an experimental challenge. Recently, polarization-analyzing gratings based on metasurface technology~\cite{Rubin2019, Rubin2021a} have enabled a new class of optical components that can diffract light into a set of diffraction orders with each of them acting as a polarization analyzer for an arbitrarily specified polarization state. By measuring the power on these diffraction orders using photo-detectors, the full-Stokes parameters ($\vec{S}$) of the incident light can be reconstructed with a pre-calibrated instrument matrix of the grating, enabling a simultaneous, passive measurement with a single optical component~\cite{Rubin2019}. Moreover, if the grating is included in the pupil plane of an imaging system, this can be extended over a field-of-view to enable polarization \emph{imaging} in the same manner~\cite{Rubin2022}.

These metasurface polarization-analyzing gratings have found applications in various field, including polarization photography~\cite{Rubin2022}, snapshot Mueller matrix imaging~\cite{Zaidi2024} and the preparation of complex polarization-entangled photon states in quantum optics~\cite{Yousef2025}. More recently, their applications have extended to compact polarization cameras for atmospheric remote sensing~\cite{Li2025} and more mature demonstrations in astronomy~\cite{Li2026}, particularly for mapping solar magnetic fields in space weather studies.

These devices are designed using a formalism known as Jones matrix Fourier optics~\cite{Rubin2019, Rubin2021a}, in which the polarization-sensitive nature of the diffraction orders can be mapped to a spatially-varying polarization transformation implemented by the metasurface grating using Fourier transform. This technique involves several simplified assumptions and is not exact with respect to Maxwell's Equations. Most notably, the Jones matrix Fourier optics formalism --- by its very construction based on the Jones calculus~\cite{Jones1941, Damask2005, Chipman2019} which only describes plane wave polarization states --- is fundamentally paraxial.

The approximations inherent in this approach suggest that more sophisticated design techniques incorporating a more rigorous treatment of electromagnetic theory hold the potential to yield better-performing devices, especially at high diffraction angles where the paraxial approximation above notably fails. In recent years, ``inverse design'' has become a widely-accepted approach in nanophotonics by optimizing device geometries through the adjoint gradient method~\cite{Molesky2018}. The adjoint method offers a physics-based approach to calculate gradients of a merit function based on electric field simulations. These gradients provide directional guidance to the gradient descent based optimizer, facilitating the convergence toward an optimal design. By using adjoint method, the gradient for each design variable can be calculated simultaneously with one forward simulation and one adjoint simulation\cite{Lalau-Keraly2013, Miller2012}. In the realm of free-space grating design, one of the earliest examples of the adjoint method is the work by Sell et al., who demonstrated a metagrating based on freeform geometries for beam deflection which achieved absolute diffraction efficiencies over 75\% for both TE and TM polarizations, outperforming traditional forward designs. By formulating tailored objective functions, inverse designs based on adjoint methods can achieve superior performance\cite{Sell2017}. However, the application of these techniques to the design of polarization-analyzing gratings to generalize the aforementioned Jones matrix Fourier optics approach has been limited. We are aware of just one recent work~\cite{Liu2025} in this area which (inspired by the original work of~\cite{Rubin2019}) designed six-order polarization analyzing gratings by optimizing for the efficiency of these gratings directly with the adjoint method.

In this work, we propose a substantially more general topology optimization framework for the design of these polarization-sensitive metasurface gratings using adjoint-based inverse design, which achieves both large grating angles and better performance metrics compared to previous methods. In particular, we optimize with regard to a polarization-sensitive objective function that explicitly considers the mathematical optimality of the grating when used as a full-Stokes polarimeter. To validate the design framework, we performed full-wave simulations of the structure and evaluated its performance metrics. Additionally, we also fabricated one of these freeform grating designs using TiO$_2$ on glass and measured its performance which (although somewhat different than the simulated result) is functional and performs roughly as expected.

This work holds promise for the design of high-performance metasurface polarization gratings in polarization-sensitive applications in optical systems of all kinds.

\section{Designing Polarization-Analyzing Metasurface Gratings}
This section begins with a description of the traditional forward method for polarization analyzing grating design. This method is constructed around the Jones matrix formalism and is therefore restricted to plane-wave polarization states; this traditional forward method is fundamentally a paraxial technique. We then introduce an alternative method, where the grating design is optimized directly through Maxwell's equations by employing the adjoint gradient method. This technique overcomes the paraxial limitation of the traditional forward method using differentiable full-wave simulations to obtain the polarization response of the grating, thereby surmounting the approximation inherent in the simpler Jones matrix technique.

\subsection{Original Method Based on Jones Matrix Fourier Optics}
\label{subsec:matrix_fourier_optics}

The design of metasurface polarization-analyzing gratings using Jones matrix Fourier optics was introduced in~\cite{Rubin2019} and detailed in its supplementary information. The method is resummarized here. We consider a two-dimensional grating operating at wavelength $\lambda$, with square repeating unit cells of side length $D$. This geometry is useful for imaging applications~\cite{Rubin2019}, although the framework can be extended to one-dimensional gratings and other unit-cell geometries depending on the application. Under normal incidence, the first-order diffraction angle $\theta_{\mathrm{grating}}$ is defined by

\begin{equation}
\label{eq:grating_angle}
    \sin\theta_{\mathrm{grating}}=\frac{\lambda}{D}
\end{equation}

Each repeating unit cell contains an $N\times N$ array of dielectric pillars with center-to-center spacing $d$, such that $D=Nd$.

Each pillar is characterized by its lateral dimensions $d_x(m,n)$ and $d_y(m,n)$ along its principal axes and its rotation angle $\theta_{m,n}$ about the $z$-axis, where $m,n=1,\ldots,N$ label the pillar positions within a unit cell. Assuming a local, lossless response, the Jones matrix of each pillar is
\begin{equation}
    \mathbf{J}_{m,n}
    =
    \mathbf{R}(\theta_{m,n})
    \begin{pmatrix}
        e^{i\phi^x_{m,n}} & 0 \\
        0 & e^{i\phi^y_{m,n}}
    \end{pmatrix}
    \mathbf{R}(-\theta_{m,n})
\end{equation}
where
\begin{equation}
    \mathbf{R}(\theta)
    =
    \begin{pmatrix}
        \cos\theta & -\sin\theta \\
        \sin\theta & \cos\theta
    \end{pmatrix}
\end{equation}
is the in-plane rotation matrix. The phase delays $\phi^x_{m,n}$ and $\phi^y_{m,n}$ describe the responses along the pillar's principal axes (the local `$x$' and `$y$' axes) and depend on $d_x(m,n)$ and $d_y(m,n)$ at the operating wavelength, with the remaining structural and material parameters fixed.

The near-field, spatially-varying Jones matrix experienced by incident light at location $(x,y)$ can be written by the summation of the Jones matrices for each individual pillar across the grating area as

\begin{equation}
    \mathbf{J}_{\mathrm{cell}}(x,y)
    =
    \sum_{m=1}^{N}\sum_{n=1}^{N}
    \mathbf{J}_{m,n}
    \operatorname{rect}\!\left(\frac{x-x_m}{d}\right)
    \operatorname{rect}\!\left(\frac{y-y_n}{d}\right)
\end{equation}
where $x_m=(m-\tfrac{1}{2})d$ and $y_n=(n-\tfrac{1}{2})d$ are the pillar-center coordinates. The function $\operatorname{rect}(u)$ equals one for $|u|<1/2$ and zero for $|u|>1/2$, so that $\mathbf{J}_{m,n}$ is assigned to a square lattice cell of side length $d$. This unit cell repeats with period $D=Nd$ along both directions.

The Jones response of diffraction order $(p,q)$ is given by the corresponding Fourier series coefficient of this periodic near-field model:
\begin{equation}
    \begin{aligned}
        \mathbf{A}_{p,q}
        ={}&
        \frac{1}{N^2}
        \operatorname{sinc}\!\left(\frac{p}{N}\right)
        \operatorname{sinc}\!\left(\frac{q}{N}\right)
        \\
        &\times
        \sum_{m=1}^{N}\sum_{n=1}^{N}
        \mathbf{J}_{m,n}
        \exp\!\left[
            -i\frac{2\pi}{N}
            \left(
                p\left(m-\frac{1}{2}\right)
                +q\left(n-\frac{1}{2}\right)
            \right)
        \right]
    \end{aligned}
\end{equation}
where $p,q\in\mathbb{Z}$ denote the diffraction-order indices and $\operatorname{sinc}(u)=\sin(\pi u)/(\pi u)$, with $\operatorname{sinc}(0)=1$. The sinc factors account for the finite extent of each square lattice cell.

To define the optimization objective, let $\mathcal{S}$ denote the set of target diffraction orders. Each order $(p,q)\in\mathcal{S}$ is assigned a normalized incident polarization state $\lvert s_{p,q}\rangle$ to be analyzed. Its normalized orthogonal state, $\lvert s_{p,q}^{\perp}\rangle$, satisfies
\begin{equation}
    \langle s_{p,q}\vert s_{p,q}\rangle
    =
    \langle s_{p,q}^{\perp}\vert s_{p,q}^{\perp}\rangle
    =1,
    \qquad
    \langle s_{p,q}^{\perp}\vert s_{p,q}\rangle=0
\end{equation}
Within the paraxial model, the powers directed into this order for these two incident states, normalized to the incident power, are
\begin{align}
    I_{\parallel}^{(p,q)}
    &=
    \langle s_{p,q}\vert
    \mathbf{A}_{p,q}^{\dagger}\mathbf{A}_{p,q}
    \vert s_{p,q}\rangle,
    \\
    I_{\perp}^{(p,q)}
    &=
    \langle s_{p,q}^{\perp}\vert
    \mathbf{A}_{p,q}^{\dagger}\mathbf{A}_{p,q}
    \vert s_{p,q}^{\perp}\rangle
\end{align}
where $\mathbf{A}_{p,q}$ is the Jones matrix response of diffraction order $(p,q)$ defined above, and $\dagger$ denotes the Hermitian conjugate. Here, $I_{\parallel}^{(p,q)}$ describes the response to the assigned polarization, whereas $I_{\perp}^{(p,q)}$ describes the response to its orthogonal polarization. The polarization contrast is defined as the difference between the powers in a given diffraction order where the incident polarization states are parallel and orthogonal respectively, and then normalized by their sum:

\begin{equation}
    \xi_{p,q}
    =
    \frac{I_{\parallel}^{(p,q)}-I_{\perp}^{(p,q)}}
         {I_{\parallel}^{(p,q)}+I_{\perp}^{(p,q)}}
\end{equation}
For a nonzero response, $I_{\parallel}^{(p,q)}+I_{\perp}^{(p,q)}>0$, the condition $\xi_{p,q}=1$ indicates complete rejection of the orthogonal polarization.

The optimization seeks to maximize the responses of the target orders to their assigned polarizations while enforcing ideal polarization contrast. Following Ref.~\cite{Rubin2019}, the merit function is
\begin{equation}
    \eta
    =
    \sum_{(p,q)\in\mathcal{S}}I_{\parallel}^{(p,q)}
\end{equation}

Equivalently, defining the loss as $\mathcal{L}_{\mathrm{forward}}=-\eta$, the optimization problem is
\begin{equation}
    \begin{aligned}
        \underset{\boldsymbol{\gamma}}{\operatorname{minimize}}
        \quad&
        \mathcal{L}_{\mathrm{forward}}(\boldsymbol{\gamma})
        =
        -\sum_{(p,q)\in\mathcal{S}}
        I_{\parallel}^{(p,q)}(\boldsymbol{\gamma}),
        \\
        \operatorname{subject\ to}\quad&
        \xi_{p,q}(\boldsymbol{\gamma})=1,
        \qquad (p,q)\in\mathcal{S}
    \end{aligned}
\end{equation}
where
\begin{equation}
    \boldsymbol{\gamma}
    =
    \left\{
        \phi^x_{m,n},\phi^y_{m,n},\theta_{m,n}
    \right\}_{m,n=1}^{N}
\end{equation}
contains the $3N^2$ design parameters. Here, again, $\phi^x_{m,n}$ and $\phi^y_{m,n}$ are the phase
delays along the two principal axes of the pillar at site $(m,n)$, and $\theta_{m,n}$ is its in-plane rotation angle. Starting from random initial values, these parameters are optimized using a gradient-based method with Lagrange multipliers to enforce the contrast constraints. The optimized phase delays are subsequently mapped to pillar dimensions using a precomputed library of pillar responses~\cite{Rubin2019}.

It is important to note that this method uses a Jones matrix formulation (which, rigorously speaking, can only describe the polarization state of plane waves) to describe the spatially-varying polarization transformation imparted by the diffraction grating unit cell; thus, this method operates within the paraxial regime. This formulation does not describe the longitudinal electric field (along $z$) that forms when plane waves interfere at high angles, which is the case for small grating periods. Moreover, the method assumes that the response of individual nanopillars can be drawn from a pre-simulated library, in which each pillar geometry is simulated using periodic boundary conditions that represent an infinite array of identical pillars. This, too, is a sort of small angle approximation that neglects coupling and cross-talk between adjacent grating elements.

While the method developed in~\cite{Rubin2019} has been used successfully in several works for a variety of applications (\textit{e.g.}, ~\cite{Rubin2019, Zaidi2024, Li2025, Li2026}), it is fundamentally a numerical abstraction that neither directly incorporates Maxwell's Equations nor optimizes based on real structures. A more complete method free of these approximations could yield higher performance devices, particularly at high grating angles where the paraxial approximation inherent in the Jones calculus is not valid.

\subsection{Alternative End-To-End Inverse Design Formalism}
\label{subsec:inverse_design_method}

\begin{figure*}[htbp] 
  \centering
  \includegraphics[width=0.8\linewidth]{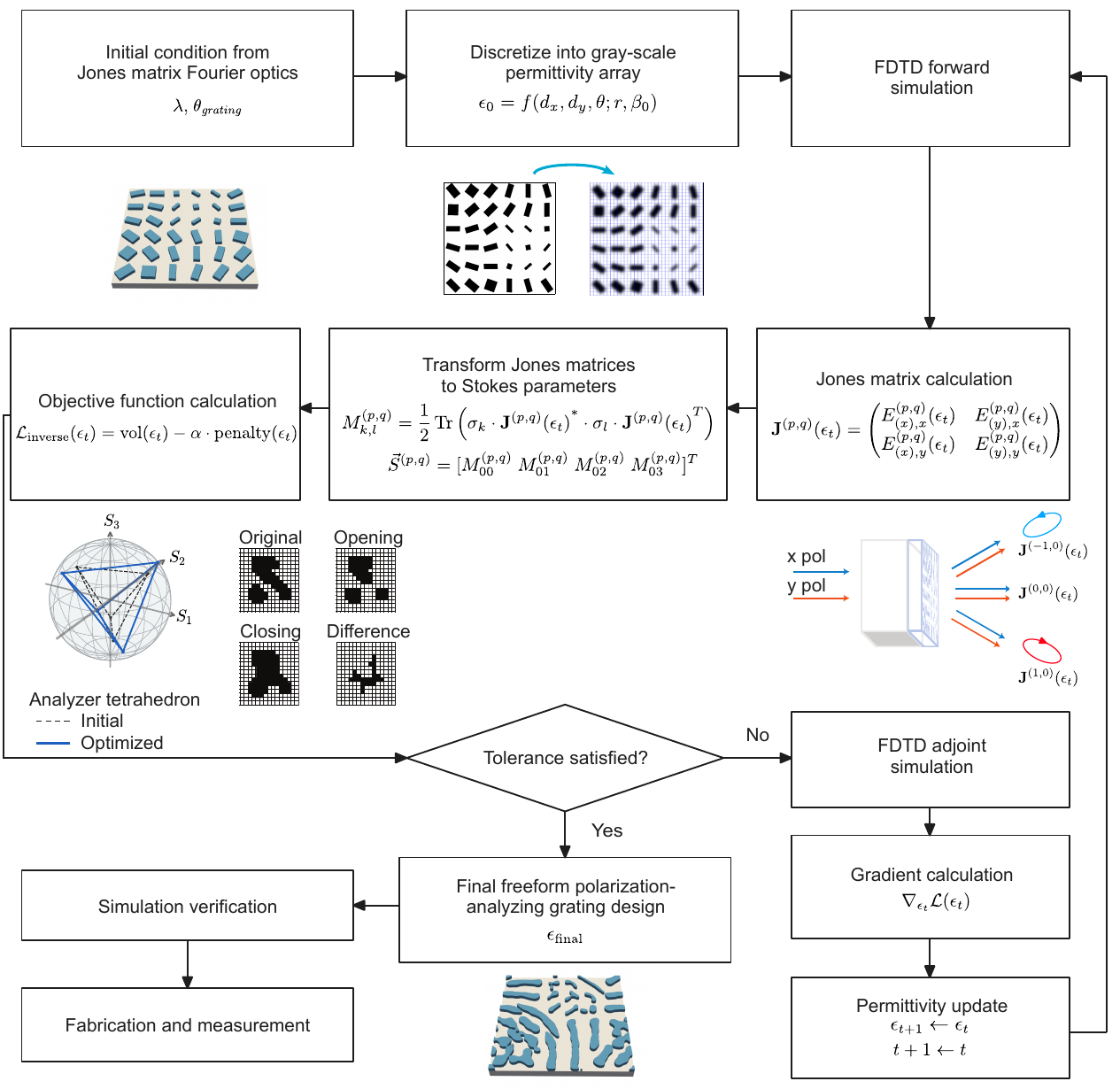} 
  \caption{\textbf{Differentiable design of polarization-analyzing metasurface gratings with full Maxwell's Equations in-the-loop}: The initial condition (top left) starts from a simpler paraxial design generated by the forward method of Sec. \ref{subsec:matrix_fourier_optics}. The methods sketched in this figure are described rigorously in Sec. \ref{subsec:inverse_design_method}.}
  \label{fig:flowchart}
\end{figure*}

To overcome these major limitations, we propose an alternative design workflow using full-wave topology-optimization which directly connects the far-field polarization response to the permittivity distribution of the structure by differentiable Finite-Difference Time-Domain (FDTD) simulations. This method (as described in this section) is sketched by the flowchart in Fig. \ref{fig:flowchart}.

\subsubsection{Initial Condition}
To accelerate the optimization and improve convergence, we first generate an initial condition using the traditional forward method described above. This yields a design of discrete pillars which are then discretized on a pixelated lattice into a permittivity distribution given by $\epsilon_0$ (the `0' here refers to optimization step 0, not to be confused with the permittivity of free space $\epsilon_0$) as

\begin{equation}
\epsilon_0 = f(d_x, d_y, \theta; r, \beta_0)
\end{equation}

where $f$ is a projection function that maps the geometric parameters of the pillar-based grating to a discretized permittivity distribution with resolution $r$ (in nm) and $\beta_0$ is the initial value of the projection parameter, which controls the sharpness of binary projection. The permittivity distribution $\epsilon_0$ is a function of 2D space: $\epsilon_0(x,y)$.

This $f$ function is also applied in the subsequent optimization process for geometry pattern construction. $f$ can be understood as a combined filtering and projection transformation where the filtering step smooths small-scale variations in the design array, thus discouraging features below the length scale set by the filter kernel while the projection step applies a tanh function, promoting a binary material distribution since the filtering introduces grayscale values near material boundaries. The projection is gradually sharpened by increasing the binary projection parameter $\beta$ as optimization proceeds. With our chosen settings, the structure exhibits clear material boundaries after approximately $65$ optimization steps, before the small-feature penalty is introduced.

\subsubsection{Grating Simulation and Order Characterization}

The grating is then optimized from this starting condition. At each iteration $t$, the polarization response of the grating with permittivity distribution $\epsilon_t$ is obtained from two separate FDTD simulations, using $x$- and $y$-polarized incident plane waves of unit amplitude as the excitation sources. The polarization response of each diffraction order $(p, q)$, as expressed by the order's Jones matrix $\mathbf{J}^{(p,q)}(\epsilon_t)$, can be extracted from these simulation results as:

\begin{equation}
\label{eq:sim_results}
    \begin{aligned}
        &\mathbf{J}^{(p,q)}(\epsilon_t)
        \begin{pmatrix} 1 \\ 0 \end{pmatrix}
        =
        \begin{pmatrix}
            E_{(x),x}^{(p,q)}(\epsilon_t) \\
            E_{(x),y}^{(p,q)}(\epsilon_t)
        \end{pmatrix},
        \\
        &\mathbf{J}^{(p,q)}(\epsilon_t)
        \begin{pmatrix} 0 \\ 1 \end{pmatrix}
        =
        \begin{pmatrix}
            E_{(y),x}^{(p,q)}(\epsilon_t) \\
            E_{(y),y}^{(p,q)}(\epsilon_t)
        \end{pmatrix}
    \end{aligned}
\end{equation}

In $E_{(\nu),\mu}^{(p,q)}$, the parenthesized subscript $\nu\in\{x,y\}$ specifies the incident polarization, while $\mu\in\{x,y\}$ specifies the output electric-field component. The right side of each expression in Eq. \ref{eq:sim_results} connotes the full vectorial electric field on each diffraction order $(p, q)$ for $x$- and $y$- polarized illumination, respectively. These fields can be computed from far-field projection, \textit{i.e.}, Fourier transformation of the vectorial electric field just above the grating surface in the FDTD simulation.

These two responses can then be combined to form the first and second columns of the order's Jones matrix, respectively, to constrain the full-polarization response of a given diffraction order as

\begin{equation} 
    \mathbf{J}^{(p,q)}(\epsilon_t)
    =
    \begin{pmatrix}
        E_{(x),x}^{(p,q)}(\epsilon_t)
        &
        E_{(y),x}^{(p,q)}(\epsilon_t)
        \\
        E_{(x),y}^{(p,q)}(\epsilon_t)
        &
        E_{(y),y}^{(p,q)}(\epsilon_t)
    \end{pmatrix}
\end{equation}

The Jones matrix of each diffraction order is further transformed into its corresponding Mueller matrix via Pauli matrix decomposition. For the $(p,q)^{\text{th}}$ diffraction order, the Mueller matrix is denoted as $\mathbf{M}^{(p,q)}$, whose $(k,l)^{\text{th}}$ element is given by

\begin{equation}
M^{(p,q)}_{k,l} = \frac{1}{2} \operatorname{Tr} 
\left( \sigma_k \cdot {\mathbf{J}^{(p,q)}(\epsilon_{t})}^{*} \cdot \sigma_l \cdot {\mathbf{J}^{(p,q)}(\epsilon_{t})}^{T} \right)
\end{equation}

where $\mathbf{J}^{(p,q)}(\epsilon_t)$ is the Jones matrix of the $(p,q)$-th diffraction order, and $\operatorname{Tr}(\cdot)$ denotes the trace operator. The matrices $\sigma_k$ ($k = 0,1,2,3$) are defined as
\begin{equation}
    \begin{aligned}
        &\sigma_0 = \begin{pmatrix} 1 & 0 \\ 0 & 1 \end{pmatrix}, \quad
        \sigma_1 = \begin{pmatrix} 1 & 0 \\ 0 & -1 \end{pmatrix}, \quad
        \\
        &\sigma_2 = \begin{pmatrix} 0 & 1 \\ 1 & 0 \end{pmatrix}, \quad
        \sigma_3 = \begin{pmatrix} 0 & -i \\ i & 0 \end{pmatrix}.
    \end{aligned}
\end{equation}

which correspond to four Pauli matrices (including the identity matrix) and are ordered according to the convention typical in polarization optics (notably different than that used in quantum mechanics).

The first row of the Mueller matrix characterizes the power transfer from the incident polarization state to the output beam and fully defines the polarization-analyzing intensity-contrast behavior of any polarization-modifying transformation. Accordingly, in this work, we extract only the first row of $\mathbf{M}^{(p,q)}$ to represent the analyzer state realized by the freeform grating on each diffraction order. Then, the analyzer polarization states of a given diffraction order with indices $(p,q)$ can be expressed as:

\begin{equation}
\begin{aligned}
    \vec{S}^{(p,q)} 
    &= [M^{(p,q)}_{00} \space M^{(p,q)}_{01} \space M^{(p,q)}_{02} \space  M^{(p,q)}_{03}]^T \quad
    \\
     &= [S^{(p,q)}_{0} \space S^{(p,q)}_{1} \space S^{(p,q)}_{2} \space  S^{(p,q)}_{3}]^T
\end{aligned}
\label{eq:order_stokes}
\end{equation}

Here, the first expression uses the first row of the order’s Mueller matrix, while the second introduces shorthand notation used below. The vector $\vec{S}^{(p,q)}$ can be interpreted as Stokes parameters at diffraction order $(p,q)$ and connote the full polarization-analyzing behavior, including efficiency, diattenuation, and the polarization state to which the order is sensitive.

\subsubsection{Merit criterion in terms of tetrahedron volume}

We next define a quantitative merit criterion for the polarization responses of the targeted diffraction orders. In this work, our goal is to design a grating device for full-Stokes polarimetry, where the diffraction orders of interest implement distinct polarization analyzers as connoted by Eq. \ref{eq:order_stokes}~\cite{Rubin2019}. By measuring the intensity directed to these orders, the polarization state incident on the grating sample can be fully determined. Each measured intensity is a linear combination of the four elements of the incident Stokes parameters and recovering all of them therefore requires at least four independent measurements, with the analyzer vectors defined in Eq. \ref{eq:order_stokes} forming an instrument matrix of rank four. The measurement and reconstruction procedures are detailed in Sec \ref{subsec: fabrication experiment} and the Supplemental Information.

While the formalism for this information retrieval is general and agnostic to the precise set of any four diffraction orders $\{(p, q)\}$ targeted, in this work we presume that gratings concentrate light into the innermost set of four diffraction orders (as in past work~\cite{Rubin2019, Li2025}), \textit{i.e.}, the first orders with indices $(p, q) \in \{(+1, 0), (-1, 0), (0, +1), (0, -1)\}$.

In addition to invertibility, reliable reconstruction requires a combination of analyzer states that minimize the amplification of measurement noise. We need to search for a group of analyzer states that are as unique as possible with high diattenuation, high efficiency, and balanced power distribution. The condition number of this measurement matrix is then used to evaluate this property, as discussed in \ref{subsec: performance metrics}.

These design considerations can be incorporated into a single merit criterion using the Poincar\'e sphere representation as depicted by the sphere inset in Fig. \ref{fig:flowchart}. From the polarization information of each targeted diffraction order described in Eq. \ref{eq:order_stokes}, we extract the last three elements and represent them as coordinates along the three axes of the Poincar\'e sphere without normalization:
\begin{equation}
    \mathbf{s}^{(p,q)}=[S_1^{(p,q)},S_2^{(p,q)},S_3^{(p,q)}]^T
\end{equation}
where the directions of the vector represents the analyzed polarization state and the radius (norm) of each vector $\mathbf{s}^{(p,q)}$ by definition can be expressed as:
\begin{equation}
   \begin{aligned}          
   \|\mathbf{s}^{(p,q)}\|
   &=S_0^{(p,q)}\mathcal{D}^{(p,q)}
   \\
   &=\sqrt{(S_1^{(p,q)})^2+(S_2^{(p,q)})^2+(S_3^{(p,q)})^2}
   \end{aligned} 
\end{equation}
with
\begin{equation}
\mathcal{D}^{(p,q)}
    =
    \frac{
        \sqrt{(S_1^{(p,q)})^2+(S_2^{(p,q)})^2+(S_3^{(p,q)})^2}
    }{S_0^{(p,q)}}
    \label{eq:diattenuation}
\end{equation}
where $\mathcal{D}^{(p,q)}$ is the diattenuation of order $(p,q)$ and $S_0^{(p,q)}$ is its average transmission efficiency over all possible input polarization states.

The four vectors correspond to the four targeted diffraction orders that form the vertices of a tetrahedron. We use its volume as an objective function to evaluate the performance of a given grating as a full-Stokes polarimeter accounting for polarization-state separation, diffraction efficiency and diattenuation (in terms of the conditioning of the aforementioned inverse problem).
As this volume increases, the utility of the grating for full-Stokes polarimetry increases --- from the worst possible case when, \textit{e.g.}, the polarization states analyzed by the four orders are identical and the volume is identically zero, to the best possible case in which the polarization states of the four orders correspond to a regular tetrahedron (\textit{i.e.}, they are maximally distinct from each other), offer equal and maximal efficiency (\textit{i.e.}, each has an efficiency of 25\% in the best case), with diattenuation of unity. This best case maximizes the volume of this tetrahedron. 

At iteration $t$, the tetrahedron volume for the permittivity distribution $\epsilon_t$ is
\begin{equation}
\label{eq:vol_condition}
\begin{aligned}
    \mathrm{vol}(\epsilon_t) ={}&
    \frac{1}{6}\Bigl|
    \bigl(\mathbf{s}^{(+1,0)}-\mathbf{s}^{(-1,0)}\bigr)
    \cdot
    \\
    &
    \Bigl[
    \bigl(\mathbf{s}^{(0,-1)}-\mathbf{s}^{(-1,0)}\bigr)
    \times
    \bigl(\mathbf{s}^{(0,+1)}-\mathbf{s}^{(-1,0)}\bigr)
    \Bigr]
    \Bigr|.
    \end{aligned}
\end{equation}
which can serves as the merit criterion for the polarization response of the current grating design. To account for fabrication constraints, we combine this volume with a feature penalty:
\begin{equation}
\label{eq:merit_criterion}
\mathcal{L}_{\mathrm{inverse}}(\epsilon)
=
\mathrm{vol}(\epsilon)
-\alpha\,\mathrm{penalty}(\epsilon)
\end{equation}

The optimized permittivity distribution $\epsilon^*(x,y)$ is obtained as
\begin{equation}
\epsilon^*=\underset{\epsilon}{\arg\max}\,\mathcal{L}_{\mathrm{inverse}}(\epsilon)
\end{equation}

Simply put, we seek to optimize the design of the grating (\textit{i.e.}, $\epsilon$) with respect to Eq. \ref{eq:vol_condition}, subject to a penalty. We introduce this penalty to optimize for robustness of the grating to errors in fabrication. Specifically, $\mathrm{penalty}(\epsilon)$ is defined as follows, which is a fabrication penalty

\begin{equation}
\text{penalty}(\epsilon_t) =
\frac{
\left\|
\bigl((\epsilon_t \oplus B)\ominus B\bigr)
-
\bigl((\epsilon_t \ominus B)\oplus B\bigr)
\right\|_{F}
}{\sqrt{N}}
\label{eq:feature penalty}
\end{equation}

where $\oplus$ and $\ominus$ denote the morphological dilation and erosion operations, respectively, $\|\cdot\|_{F}$ denotes the Frobenius norm, $B$ is the structuring element defining the minimum feature size (kernel for morphological operations) and $N$ is the total number of pixels in the design region.
Morphological closing, $\bigl((\epsilon_t \oplus B)\ominus B\bigr)$, consists of dilation followed by erosion and tends to fill small gaps. Morphological opening, $\bigl((\epsilon_t \ominus B)\oplus B\bigr)$, consists of erosion followed by dilation and tends to remove small solid features.
The objective function illustrated in Fig.\ref{fig:flowchart} shows how Eq. \ref{eq:feature penalty} calculates the  differences in material distribution $\epsilon_t$ after morphological closing and opening. 
This penalty first calculates the Frobenius norm of the difference between the opened and closed permittivity distribution and this value is divided by the square root of the pixel number in the permittivity array $\epsilon_t$ (equivalent to the root-mean-square difference). This metric provides a measure of the sensitivity of the design to fabrication-induced geometric variations. The penalty depends on both the design geometry and the kernel size, with the value smaller than one. The volume of an ideal regular tetrahedron inscribed in a sphere with radius of 0.25 is approximately $0.008$, which provides a reference scale for the polarization objective. We therefore choose $\alpha=0.001$ in Eq.~\ref{eq:merit_criterion} to limit the contribution of the feature penalty while retaining its influence on the optimization. To allow the optimizer to first learn the material distribution of a high polarization performance, we introduce the feature penalty only after the first $75$ steps.

The gradient of the objective function with respect to the permittivity distribution is computed using the adjoint method. For each diffraction order and polarization state, a forward simulation is first performed to obtain the electromagnetic fields. An adjoint simulation is then constructed by defining adjoint sources based on the sensitivity of the objective function with respect to the output fields. The gradient related to objective function is determined by the chain rule and handled automatically by chain rule via Jax, via the differentiable FDTD solver implemented in Tidy3D (FlexCompute, Inc.). Conveniently, since all the blocks of Fig. \ref{fig:flowchart} are natively differentiable (including the finite-difference solver), the construction of the adjoint sources is handled automatically, and no manual computations of analytical expressions of the gradients are required.

\section{Results and Analysis}
We now apply the method detailed above to the inverse design of polarization-analyzing metasurface gratings. In the numerical and experimental examples that follow, we assume the use of TiO$_2$ in the visible at $\lambda=870$ nm.  In this work, all designs have a uniform structure height of $700$ nm across the entire design area in both simulations and fabricated devices. These material and pillar height choices are not fundamental, but TiO$_2$ has been widely employed for these devices in the past for visible wavelengths and 870 nm is a wavelength that -- by virtue of being redder than the typical visible wavelengths for which TiO$_2$ is used -- permits slightly larger feature sizes and thus relaxes fabrication tolerances.

Below, we first define figures-of-merit for understanding the performance of polarization-analyzing gratings once designed. We then examine a specific grating as a test-case of the advantage offered by this method, then show that this advantage holds over a wide range of grating periods --- with a significant advantage offered for small periods/large angles. We conclude with an experimental realization of one of these high angle gratings.

\subsection{Performance Metrics}
\label{subsec: performance metrics}
Three parameters are considered when evaluating these gratings designs: the efficiency and the diattenuation of the orders, along with how distinct the polarization states analyzed are. These three criteria together inform how effectively the grating can be used as a full-Stokes polarimeter.

For each order, the efficiency is a unitless value and is the first Stokes parameter,  $S^{(p,q)}_0$ (which is also the upper-left element of the order's Mueller matrix $M^{(p,q)}_{00}$). This element represents the mean power transmitted over all possible input polarization states, and would be the fraction of input power directed to the order if unpolarized light were incident on the grating.

The diattenuation $\mathcal{D}^{(p,q)}$ is given in Eq.~\eqref{eq:diattenuation}. It reflects the polarization contrast of the diffraction order, which is also equal to the degree of polarization of light that would emerge on the diffraction order if the device were illuminated by unpolarized light.

While the gratings are optimized using the tetrahedron volume metric defined above, we instead consider the \emph{condition number} of the reconstruction matrix implied by the grating as a polarimeter to assess the grating design quality after optimization. The condition number connotes equivalent information as the volume metric above, but is more widespread in polarimetry and is simpler to interpret.

Specifically, the four polarization states analyzed by the grating can be written as the four rows of an instrument matrix for the grating as

\begin{equation}
    A_{\mathrm{grating}} = \begin{pmatrix}
        (\vec{S}^{(1,0)})^T\\
        (\vec{S}^{(-1,0)})^T\\
        (\vec{S}^{(0,1)})^T\\
        (\vec{S}^{(0,-1)})^T
    \end{pmatrix}
\end{equation}

where each row is the unnormalized Stokes parameters characteristic of each diffraction order

For a full-rank analyzer matrix, the condition number of the freeform grating $\kappa(A_{\mathrm{grating}})$ can be expressed as:
\begin{equation}
    \kappa(A_{\mathrm{grating}})
    =
    \left\lVert A_{\mathrm{grating}}^{-1}\right\rVert
    \left\lVert A_{\mathrm{grating}}\right\rVert
    =
    \frac{\sigma_{\max}}{\sigma_{\min}}
\end{equation}
Here $\left\lVert \cdot \right\rVert$ denotes the $L_2$ norm and $\sigma_{\max}$ and $\sigma_{\min}$ represent the largest and the smallest singular value respectively.

The theoretical best-case where the four analyzer states form a regular tetrahedron with equal efficiencies and unit diattenuations on the Poincar\'e sphere yields a lower limit for $\kappa(A_{\mathrm{grating}})$ of $\sqrt{3}$. A \emph{smaller} condition number indicates less sensitivity to measurement noise and represents a more desirable grating performance.

\subsection{Simulation Study of Inversely Designed Gratings}
\begin{figure*}[htbp]
    \centering
    \includegraphics[width=0.75\textwidth]{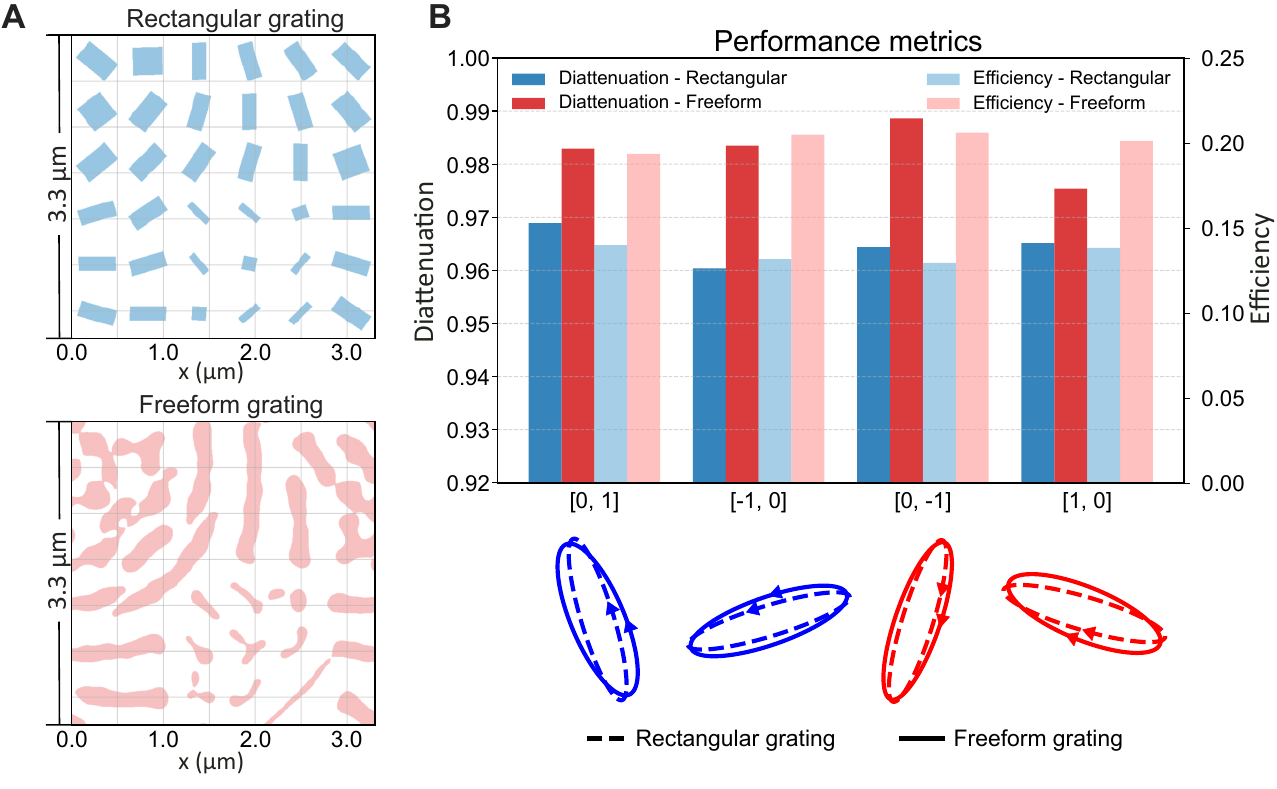} 
    \caption{\textbf{Performance comparison of a freeform grating and a rectangular pillar-based grating at a diffraction angle of $15.10^{\circ}$.}
    \textbf{A}. Unit cell designs for a rectangular pillar-based grating and a freeform grating both designed at a diffraction angle of $15^{\circ}$ and working wavelength of $870$ nm, with a uniform structure height of $700$ nm. The top design (``rectangular grating'', blue) is generated using the Jones matrix Fourier optics forward design technique of Sec. \ref{subsec:matrix_fourier_optics}, while the bottom design (``freeform grating'', red) results from the proposed full-wave inverse design method (using the top design as an initial condition). Both gratings have the same period of $3.3$ \textmu m.
    \textbf{B}. Performance metrics and polarization ellipses for the grating designs shown in A. In the bar chart, dark and light blue is diattenuation and efficiency, respectively, for the rectangular  grating, while dark and light red depict corresponding  metrics for the freeform grating. The x-axis labels denote the diffraction orders of interest. 
    Below, the polarization ellipses for the polarization states analyzed by the targeted diffraction orders are shown, with blue and red here indicating left- and right-handed polarization, respectively. Dotted and solid lines correspond to the polarization states analyzed by the rectangular and freeform gratings, respectively. 
    At the same diffraction angle, the proposed inverse technique outperforms the previous forward design method with average improvements of over $4\%$ (efficiency) and $0.01$ (diattenuation). The polarization ellipses also shift during the optimization process. 
    }
    \label{fig:N6}
\end{figure*}

We now apply the method developed theoretically in Sec. \ref{subsec:inverse_design_method} to the design of real polarization-analyzing gratings. As we show below, this method provides improvements in performance at all grating angles, but at high angles succeeds where the methods of Jones matrix Fourier optics (Sec. \ref{subsec:matrix_fourier_optics}) fail.

Fig. \ref{fig:N6} shows a first example (as stated before, using TiO$_2$ features for operation at $\lambda=$ 870 nm). In this case (somewhat arbitrarily) the first-order grating has a selected period of $3.3$ \textmu m, which implies (by Eq. \ref{eq:grating_angle}) a deflection angle at $15.10^\circ$. Fig. \ref{fig:N6}A (top, in blue) depicts the design of a grating unit cell, designed using the Jones matrix Fourier optics methods of past work (``rectangular grating''). This is then used as the initial condition $\epsilon_0$ for the optimization of an inverse-designed freeform grating in the method of Fig. \ref{fig:flowchart}. Fig. \ref{fig:N6}A (bottom, in red) shows the grating that results after 140 iterations of this procedure, yielding a unit cell with some similarities in appearance but significantly changed.

Fig. \ref{fig:N6}B displays the performance of the two gratings (forward-only and inverse designed) with regards to the metrics defined above. In the bar chart at the top of Fig. \ref{fig:N6}B, the efficiency and diattenuation of all four inner grating orders are compared for both grating designs. Blue quantities represent the forward-designed grating, while red quantities represent the inverse-designed grating. With regard to both efficiency and diattenuation, the inverse-designed grating performs better than the forward-designed one, with the diattenuation of all orders increasing by several percent, and the overall efficiency of the grating increasing from roughly 50\% to nearly 80\%.

The bottom of Fig. \ref{fig:N6}B depicts the polarization states analyzed by the four diffraction orders for both the forward-designed grating and the freeform grating. Here red and blue depict right- and left-handed polarization states, respectively. The ellipses for the forward-designed and freeform gratings are shown with dashed and solid lines, respectively. These change somewhat during the optimization (as allowed by the optimizer) in a way that improves the conditioning of the grating's instrument matrix, as we describe below.

\begin{figure*}[htbp]
    \centering
    \includegraphics[width=0.75\textwidth]{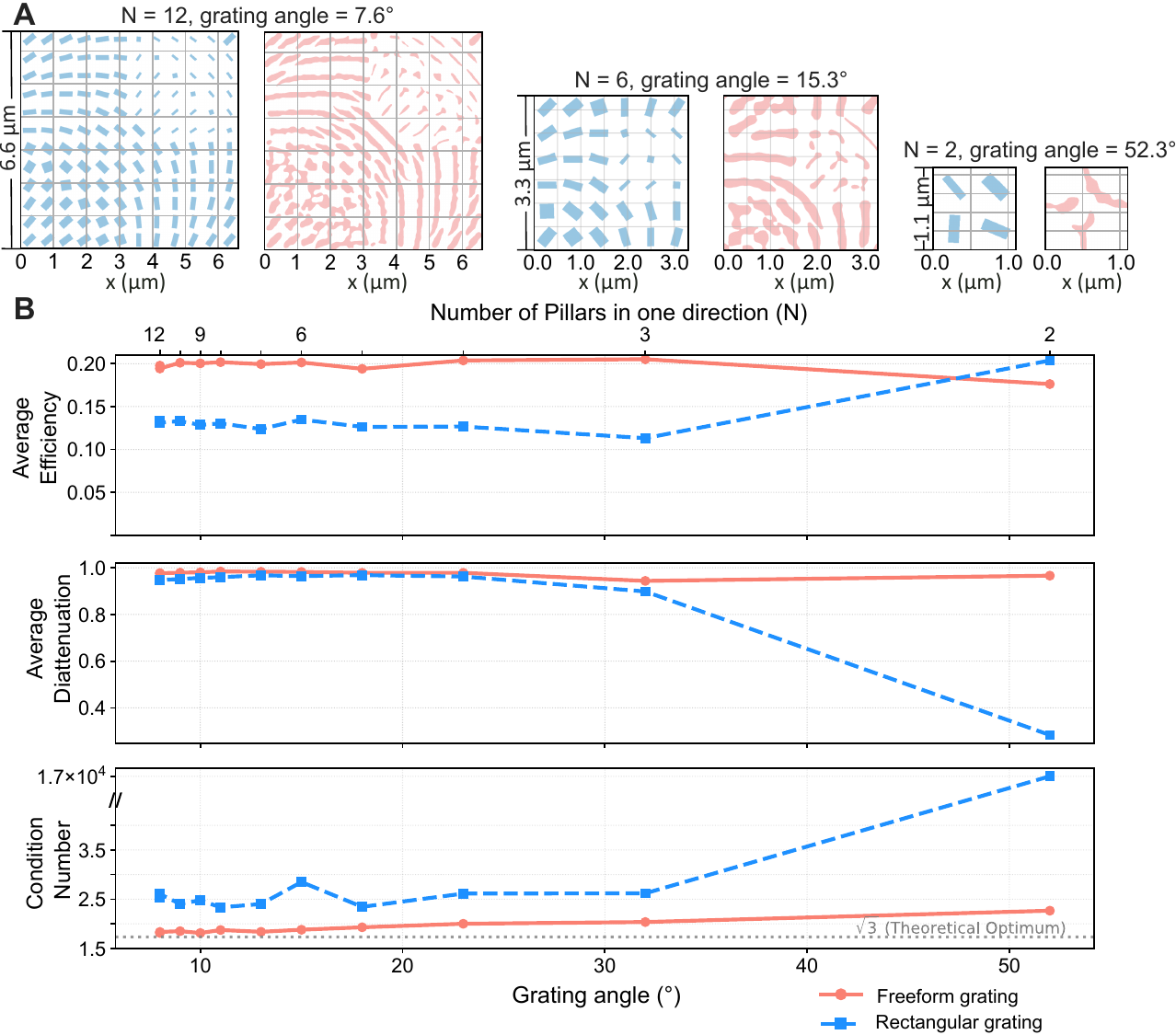} 
    \caption{\textbf{Performance of inverse designed polarization gratings for a range of grating angles}: \textbf{A}. Grating unit cell designs for a variety of grating periods (or equivalently, grating angles) for the Jones matrix Fourier optics forward design technique of Sec. \ref{subsec:matrix_fourier_optics} (blue) and the full-wave inverse design technique of this work (red). The gratings are parameterized by their unit cell period (equivalent to the number of rectangular pillars $N$ of one side at a fixed pillar-to-pillar spacing $d$). \textbf{B}. Evaluation of the grating designs for forward (blue curves) and inverse (red curves) methods. These are evaluated with respect to the average efficiency (top) and diattenuation (middle) of the inner targeted diffraction orders, as well as the conditioning of the four orders of the grating if used as a full-Stokes polarimeter (condition number, bottom). Across a wide range of angles (bottom axis), the inverse technique outperforms the previous forward design technique significantly, most notably at the highest angle studied for which the forward design technique essentially fails, while the inverse technique achieves results approaching the theoretical optimum condition number of $\sqrt{3}$ (displayed as a dotted line lower bound in the lower-most plot).}
    \label{fig:3+3}   
\end{figure*}

The one-off design of Fig. \ref{fig:N6} shows that notable performance advantages can be gained from these inverse design techniques. Next, we show that this advantage holds for a variety of grating angles. In Fig. \ref{fig:3+3}, we examine the performance of the inverse design technique by varying $N$, the number of pillars along each side of the forward-designed $N\times N$ unit cell, which changes the grating period $D=Nd$ and hence the diffraction angle according to Eq. \ref{eq:grating_angle}, with the pillar spacing fixed at $d =550$ nm. Fig. \ref{fig:3+3}A depicts three gratings designed for three different angles (with three different values of $N$). Each example shows a forward-designed grating (blue) as well as a freeform grating (red) unit cell designed by the methods of this work.

Fig. \ref{fig:3+3}B shows the performance of gratings designed by varying $N$ in discrete steps: plots depict the average efficiency and mean diattenuation of these orders, along with the condition number of $A_{\text{grating}}$ as a function of $N$ (top axis) or --- equivalently --- first-order grating angle (bottom axis). Gratings are designed and analyzed from $N=2$ (the minimum-sized grating for this choice of $d$ that still has $D>\lambda$ with propagating orders in air) to $N=12$ at which point the grating angle is $\sim7^\circ$. (We note that the upper axis of the plots in Fig. \ref{fig:3+3} may appear to be missing a tick mark, only because $N=11$ and $12$ are so close in grating angle).

It is observed from Fig. \ref{fig:3+3}B that the gratings refined by the inverse design techniques of this work (red curves) outperform their forward-designed counterparts (blue curves) for all grating unit cell periods studied here. For lower grating angles below about $30^\circ$, the forward-designs are functional for full-Stokes polarimetry. However, the inverse design process substantially improves these across-the-board --- per order efficiencies increase by more than 5\%. Diattenuation values also increase and the condition number of the grating, which reflects these improvements and the differentiation of the analyzed polarization states, are brought much closer to the theoretical best-case of $\sqrt{3}$.

At higher angles, however, the inverse design technique offers more than simple increases in performance. For the case where the foward-designed grating unit cell consists of $2\times2$ dielectric pillars, the paraxial limit underlying the Jones matrix Fourier optics technique is challenged (the grating angle is over $50^\circ$ and there are too few free-parameters in the grating unit cell to achieve function as a full-Stokes polarization analyzer). Though the efficiency of the orders increases, the diattenuation drops to below 40\% and --- as evidenced by the condition number's jump by orders-of-magnitude to $10^4$ --- the polarization analysis implemented by the grating's four orders is extremely ill-conditioned. This grating design would not be realistically usable as a full-Stokes polarization analyzer.

However, a grating designed using inverse techniques surmounts this limitation and successfully produces a design with high efficiency (each order averages 16.8\%), high average diattenuation (above 94\%), and a conditioning near the theoretical optimum of $\sqrt{3}$. At sufficiently high grating angles, the inverse design technique presented here not only \emph{improves} designs generated by more traditional means, but enables them where the existing state-of-the-art method fails.

We note that this high-angle regime is particularly important for practical applications. In polarization cameras reliant on these metasurface polarization-analyzing gratings, the grating angle sets the maximum possible field-of-view (FOV), and for any FOV (large or small) the ability to realize a high grating angle permits the fabrication of gratings with smaller aperture if relay optics are used (by conservation of \'etendue)~\cite{Rubin2022}.

In the supplemental information, the performance of some of the grating designs provided in Fig. \ref{fig:3+3}B are simulated over wide spectral ranges to examine their performance off of the design wavelength of 870 nm.

\subsection{Preliminary Experimental Demonstration}
\label{subsec: fabrication experiment}
\begin{figure*}[htbp]
    \centering
    \includegraphics[width=0.75\textwidth]{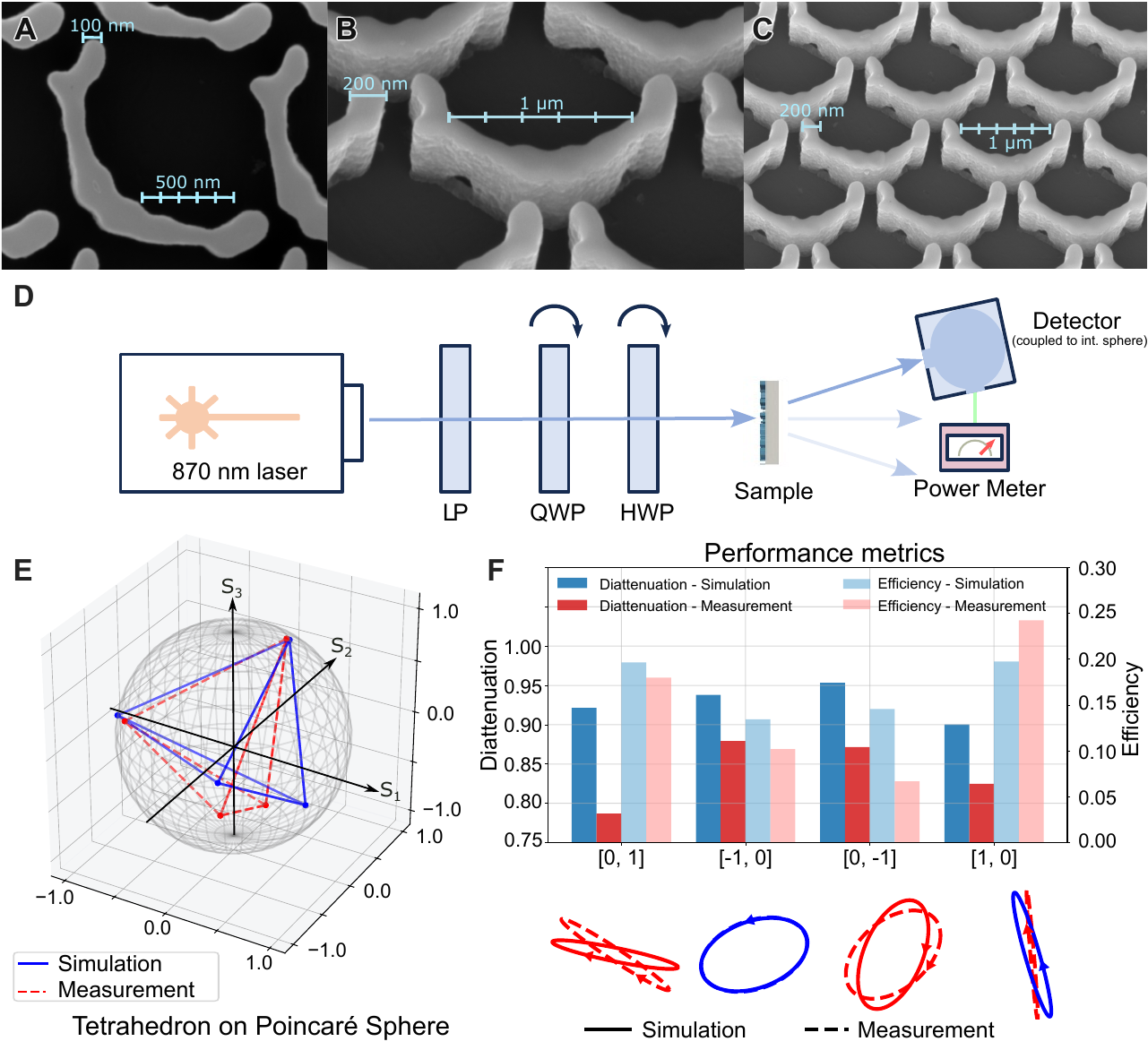} 
    \caption{\textbf{Experimental Investigation of Inverse-Designed Polarization-Analyzing Grating}
    \textbf{A-C}. Scanning electron microscope (SEM) images of the fabricated inverse-designed $\mathrm{TiO_2}$ grating sample for $870$nm with a grating angle of $52.27^{\circ}$. The grating has a period of $1.1$ \textmu m with a uniform height of $700$ nm. 
    \textbf{A}. Top view of the unit cell corresponding to the design shown in the left panel of Fig.\ref{fig:3+3}A. 
    \textbf{B}. Oblique view of the sample in A at $30^{\circ}$. 
    \textbf{C}. Oblique view of multiple unit cells at $30^{\circ}$.
    \textbf{D}. Experimental setup for polarization characterization of the fabricated sample. The laser source is followed by a fixed Linear polarizer (LP), a rotating quarter-wave plate (QWP) and a rotating half-wave plate (HWP) to generate the incident polarization states. Light on the targeted diffraction order is measured by an integrating sphere-coupled power meter.
    \textbf{E}. Tetrahedra constructed from the diattenuation-normalized Stokes parameters of the targeted orders. The solid blue tetrahedron is obtained from FDTD simulations while the red dotted line is from the sample measurements. The measured tetrahedron is and distorted rotated relative to the simulated tetrahedron. \textbf{F}. Performance metrics and polarization ellipses for the grating designs shown in A. 
    The Upper plot evaluates the performance metrics: dark and light blue bars represent the simulated diattenuation and efficiency, respectively; dark and light red bars represent the corresponding measured values. The x-axis labels denote the diffraction orders of interest. The lower plot shows the polarization ellipses for the targeted diffraction orders, with blue and red indicating left- and right-handed polarization, respectively. Dotted and solid lines denote the simulated and measured polarization ellipses. The measured diattenuation and efficiency differ from the simulated values, and the polarization ellipses exhibit changes in orientation. Nevertheless, the minimum measured efficiency and diattenuation across the targeted orders are $0.067$ and $0.78$, respectively. These results together with the tetrahedra shown in E show that under fabrication variations, the fabricated grating retains polarization selectivity at a diffraction angle of $52.27^\circ$.}
    \label{fig:SEM}
\end{figure*}

We present an experimental realization of one of these inverse-designed gratings. This demonstration illustrates what is perhaps the principal drawback of these techniques --- that the advantages offered in theory by freeform designs can be difficult to experimentally realize given the complexity of the shapes involved, even if fabrication constraints are considered in the optimization process (as has been done here).

\subsubsection{Fabrication}

We fabricated the $N=2$ design of Fig. \ref{fig:3+3}B. This design was chosen for two reasons: First, its large diffraction angle allowed us to experimentally evaluate the proposed method in the regime where the conventional approach performs poorly. Second, the unit cell size limited the design space, yielding a geometry that was easier to fabricate than other freeform designs considered with fewer small features.

The grating is fabricated in TiO$_2$ on glass using a damascene-like process (as reported by~\cite{Devlin2016}). The fabrication process involves patterning a positive electron beam resist with the grating pattern by electron beam lithography, filling in the resulting holes with TiO$_2$ by atomic layer deposition (ALD), followed by a reactive ion etch of the top layer of TiO$_2$ and a chemical strip of the resist to obtain the final nanostructure. As we discuss below, the designs produced by free-form inverse design present particular challenges to this bottom-up fabrication process. Scanning electron microscope (SEM) images of of the fabricated sample are shown in Fig. \ref{fig:SEM}A-C, both top-down (A) and oblique views (B and C).

\subsubsection{Polarimetric Characterization}

We then measure the grating, performing full-Stokes polarimetry to determine the Stokes parameters characteristic of each diffraction order (cf. Eq. \ref{eq:order_stokes}). The experimental setup is illustrated in Fig. \ref{fig:SEM}D. The sample is illuminated by continuous-wave light at $870$ nm derived from a tunable Ti:sapphire laser. A fixed linear polarizer (LP), followed by a quarter-wave plate (QWP) and a half-wave plate (HWP), vary the laser's polarization state. The orientations of the QWP and HWP were adjusted to generate different known polarization states for characterization. After interaction with the grating sample, the light was spatially separated into different diffraction orders. A sensor (coupled to an integrating sphere) is placed at a selected diffraction order to record the corresponding optical intensity in response to $n$ different input polarization states. Each of the four inner orders of the grating is measured in this way.

In measuring a given grating order $(p, q)$, let $\psi_j$ and $\phi_j$ denote the angular orientations of the fast axes of the QWP and HWP, respectively (defined relative to the fixed linear polarizer), for the $j^{\text{th}}$ measurement ($j=1,\dots,n$). The laser polarization state with Stokes parameters $\vec{S}_{\text{out}}^j$ generated by the LP, QWP and HWP based on this configuration can be expressed as

\begin{equation}
\vec{S}_{\text{out}}^j =M_{\mathrm{HWP}}(\phi_j)\,M_{\mathrm{QWP}}(\psi_j)\vec S_{\mathrm{in}}
\end{equation}

Here, $M_{\mathrm{HWP}}(\phi_j)$ and $M_{\mathrm{QWP}}(\psi_j)$ are the corresponding Mueller matrices of the HWP and QWP, respectively, and $\vec S_{\mathrm{in}}$ is the polarization state produced by passing the laser through the linear polarizer.

For each polarization state $\vec{S}_{\text{out}}^j$, the power directed to diffraction order $(p, q)$ can be measured, which we denote as $I_j$. These can be stacked into an $n$ length vector denoted as $\vec{I}$. Finally, $\vec{I}$ can be defined by dot product between the characteristic Stokes parameters of the diffraction order $\vec{S}^{(p,q)}$ and the input polarization state for each measurement $j$. This can be written compactly as

\begin{equation}
\vec{I}=A\vec S^{(p,q)}
\end{equation}

where $A$ is a $n\times4$ matrix with the polarization states input to the grating as its rows as

\begin{equation}
A=
\begin{bmatrix}
(\vec{S}_{\text{out}}^{1})^T\\
(\vec{S}_{\text{out}}^{2})^T\\
\vdots\\
(\vec{S}_{\text{out}}^{n})^T
\end{bmatrix}
\end{equation}

The Stokes parameters characterizing order $(p,q)$ can then be reconstructed as by a least-squares fit as

\begin{equation}
\vec S_{\mathrm{measured}}^{(p,q)}
=\arg\min_{\vec S^{(p,q)}}\left\|A\vec S^{(p,q)}- \vec{I} \right\|_2^2
\end{equation}

\subsubsection{Performance Results and Commentary}
\label{subsec: device measurement}
The results obtained from the fabricated sample are shown in Fig. \ref{fig:SEM}E and F. Fig. \ref{fig:SEM}E and F depict the polarization states analyzed by the selected diffraction orders, as well as their diattenuation and efficiencies. Some deviations are observed between the simulated and realized polarization analyzer states (Fig. \ref{fig:SEM}E); these states, however, retain a tetrahedral distribution on the Poincaré sphere, while exhibiting an overall rotation relative to the simulated states. This preserved (somewhat regular) tetrahedral geometry indicates that the fabricated device maintains the polarization-analyzing functionality suitable  for polarimetry.

The measured efficiency is not distributed uniformly across the targeted diffraction orders, with a minimum efficiency of approximately $0.07$. The orders, however, exhibit reasonable diattenuation, with the minimum diattenuation among the targeted orders at approximately $0.78$. Given the fabrication deviations between the designed geometry and the structures observed in the SEM images, the differences between simulation and experiment are consistent with the expected sensitivity of the optical response to geometric errors.

Overall, this sample works in a recognizable fashion, with some notable reduction in performance versus design. We note that these results are limited at present by the ability to fabricate the designs presented here with high fidelity. As we discuss below (Appendix \ref{app:fab_difficulties}), these patterns are challenging to fabricate for a variety of reasons --- both in terms of the small size of transverse features required and the high aspect ratio (with respect to the height of the structures) these then require. This is observed in Fig. \ref{fig:SEM}B, where ``bridges'' are visible at the highest aspect ratio pinch points of the structure where the (ideally) conformal TiO$_2$ growth by ALD could not reach the substrate.

In the supplemental information, we examine the effect of size defects on this pattern, as well as deviations between the sample fabricated in Fig. \ref{fig:SEM} and the theoretical design. Simulations are presented that show that this reduction in performance is expected, given the pattern that was fabricated and observed in SEM.

\section{Conclusion}

In this work, we developed a general, differentiable design framework for polarization-analyzing gratings. By moving beyond the paraxial approximation inherent in past design strategies based on Jones matrix Fourier optics, the proposed method improves the performance of metasurface polarization-analyzing gratings, particularly at large grating deflection angles. We demonstrated the utility of this method through designs which in simulation consistently outperform their forward-designed counterparts, and provided a preliminary experimental demonstration of one of these which operates at high angle and shows the desired polarization-analyzing function.

The ability to analyze multiple polarization states simultaneously makes these gratings promising for snapshot polarimetric imaging and Mueller microscopy, where polarization contrast can provide information about sample anisotropy and structure. Maintaining polarization selectivity at large diffraction angles also offers greater flexibility in separating and routing the analysis channels, potentially enabling compact full-Stokes detection modules integrated into imaging and microscopy systems with high fields-of-view.

This framework provides a flexible approach to designing polarization-sensitive optical elements for compact polarization-sensitive optical systems.

\section{Acknowledgments}

This work was supported by the Earth Science Technology Office of the National Aeronautics and Space Administration (NASA) through the Instrument Incubator Program, Grant No. 80NSSC25K7306.

\appendix
\section{Fabrication Challenges}
\label{app:fab_difficulties}

The principal drawback of this work (and, perhaps, freeform design more generally in photonics) is the difficulty of fabricating the resulting patterns, even when (as in this work) fabrication robustness considerations are explicitly included in the optimization.

There are two challenges here, one that is general and one that is specific to the fabrication method used in this work. A general challenge is complex, often small shapes that are required by the patterns that result from the inverse design procedure. Considerable lithographic optimization is required to realize these shapes, which becomes even more challenging in attempts to mass produce such designs by optical lithography.

Specific to this work (and most visible-frequency metasurface elements using TiO$_2$) is the fact that the patterns start as holes in resist, which are then filled in conformally on the bottom and sides by atomic layer deposition (ALD). This ``bottom up'' fabrication (as opposed to ``top down'' procedures in which the features would be etched into a thin film with a hard mask) presents several unique challenges. One of these is that very high aspect ratio features may ``pinch off'' before being completely filled. This is visible at the highest aspect ratio points in the SEM of Fig. \ref{fig:SEM}B, where a bridge is visible between the lower aspect ratio areas on either side of the ``neck'' of the feature in Fig. \ref{fig:SEM}A. It can be seen at this point that the TiO$_2$ failed to deposit down to the bottom of the resist hole, likely because the top of the high aspect ratio trench was closed off by TiO$_2$ deposition on the sidewalls. To some extent this could be addressed by further ALD process optimization, but this represents a general challenge in nanofabrication.

A separate but related issue is that in this fabrication procedure, close-by structures (once exposed and developed) necessitate tall, thin resist profiles which may be unstable and fall, resulting (after ALD and etching) in a distorted version of the pattern.

\begin{figure*}[htbp]
    \centering
    \includegraphics[width=0.75\textwidth]{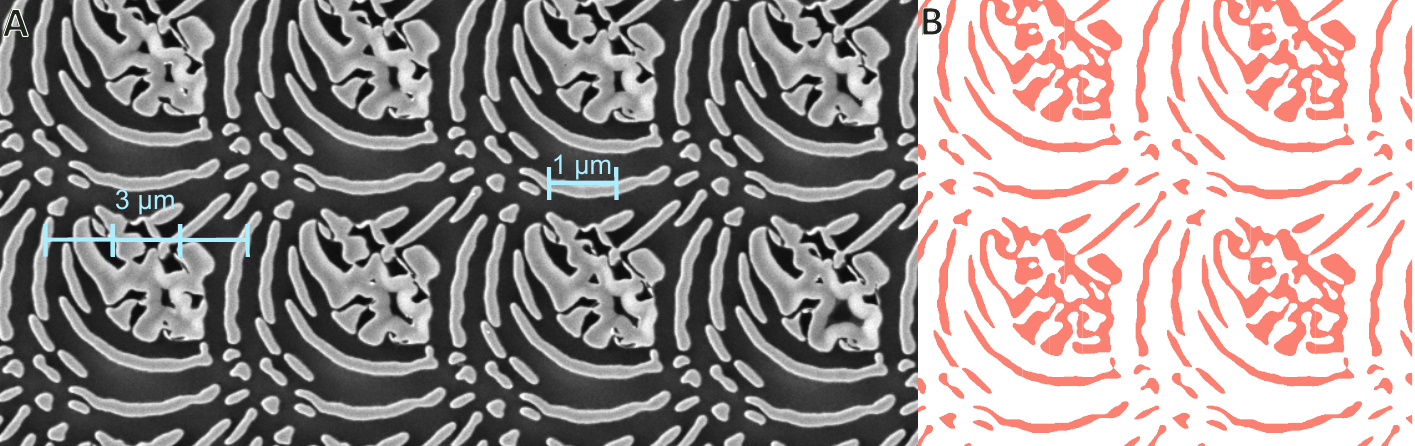} 
    \caption{\textbf{SEM images of an alternative freeform polarization-analyzing metasurface grating}: 
    \textbf{A}. SEM images of the fabricated $\mathrm{TiO_2}$ grating with a period of $3.3\mu m$.
    \textbf{B}. Corresponding geometry from the original design.}
    \label{fig:appendix_sem}
\end{figure*}

To showcase all of these issues, we present a separate high deflection angle design which was fabricated for this project but did not optically characterize. The design has a working wavelength of $870nm$ and a grating period of $3.3\mu m$. Fig. \ref{fig:appendix_sem}A shows an SEM image of the fabricated sample, while and Fig. \ref{fig:appendix_sem}B shows the initial design of the grating. Both are displayed on the same scale for comparison.

Compared with the main design discussed in this work in Fig. \ref{fig:SEM}, this design exhibits a  far more complex geometry which significantly increases the difficulty of fabrication. Notable differences between the fabricated (A) and desired (B) versions of the grating are notable by eye. In particular, the fine features in teh central region of the grating even differ substantially from unit cell to unit cell. This is particularly true in the narrow, thin regions of the grating which during fabrication would have necessitated very thin, high aspect ratio resist features to produce that likely distorted in an unpredictable way, resulting in notable differences between periodic units.

These results highlight the importance of fabrication robustness when designing complex nanophotonic structures. While highly intricate geometries may provide additional degrees of freedom in simulation, their practical performance can be strongly limited by fabrication tolerances.


%


\begin{thebibliography}{17}%
\makeatletter
\providecommand \@ifxundefined [1]{%
 \@ifx{#1\undefined}
}%
\providecommand \@ifnum [1]{%
 \ifnum #1\expandafter \@firstoftwo
 \else \expandafter \@secondoftwo
 \fi
}%
\providecommand \@ifx [1]{%
 \ifx #1\expandafter \@firstoftwo
 \else \expandafter \@secondoftwo
 \fi
}%
\providecommand \natexlab [1]{#1}%
\providecommand \enquote  [1]{``#1''}%
\providecommand \bibnamefont  [1]{#1}%
\providecommand \bibfnamefont [1]{#1}%
\providecommand \citenamefont [1]{#1}%
\providecommand \href@noop [0]{\@secondoftwo}%
\providecommand \href [0]{\begingroup \@sanitize@url \@href}%
\providecommand \@href[1]{\@@startlink{#1}\@@href}%
\providecommand \@@href[1]{\endgroup#1\@@endlink}%
\providecommand \@sanitize@url [0]{\catcode `\\12\catcode `\$12\catcode `\&12\catcode `\#12\catcode `\^12\catcode `\_12\catcode `\%12\relax}%
\providecommand \@@startlink[1]{}%
\providecommand \@@endlink[0]{}%
\providecommand \url  [0]{\begingroup\@sanitize@url \@url }%
\providecommand \@url [1]{\endgroup\@href {#1}{\urlprefix }}%
\providecommand \urlprefix  [0]{URL }%
\providecommand \Eprint [0]{\href }%
\providecommand \doibase [0]{https://doi.org/}%
\providecommand \selectlanguage [0]{\@gobble}%
\providecommand \bibinfo  [0]{\@secondoftwo}%
\providecommand \bibfield  [0]{\@secondoftwo}%
\providecommand \translation [1]{[#1]}%
\providecommand \BibitemOpen [0]{}%
\providecommand \bibitemStop [0]{}%
\providecommand \bibitemNoStop [0]{.\EOS\space}%
\providecommand \EOS [0]{\spacefactor3000\relax}%
\providecommand \BibitemShut  [1]{\csname bibitem#1\endcsname}%
\let\auto@bib@innerbib\@empty
\bibitem [{\citenamefont {Tyo}\ \emph {et~al.}(2006)\citenamefont {Tyo}, \citenamefont {Goldstein}, \citenamefont {Chenault},\ and\ \citenamefont {Shaw}}]{Tyo2006}%
  \BibitemOpen
  \bibfield  {author} {\bibinfo {author} {\bibfnamefont {J.~S.}\ \bibnamefont {Tyo}}, \bibinfo {author} {\bibfnamefont {D.~L.}\ \bibnamefont {Goldstein}}, \bibinfo {author} {\bibfnamefont {D.~B.}\ \bibnamefont {Chenault}},\ and\ \bibinfo {author} {\bibfnamefont {J.~A.}\ \bibnamefont {Shaw}},\ }\href {https://doi.org/10.1364/AO.45.005453} {\bibfield  {journal} {\bibinfo  {journal} {Applied Optics}\ }\textbf {\bibinfo {volume} {45}},\ \bibinfo {pages} {5453} (\bibinfo {year} {2006})}\BibitemShut {NoStop}%
\bibitem [{\citenamefont {Rubin}\ \emph {et~al.}(2019)\citenamefont {Rubin}, \citenamefont {D'Aversa}, \citenamefont {Chevalier}, \citenamefont {Shi}, \citenamefont {Chen},\ and\ \citenamefont {Capasso}}]{Rubin2019}%
  \BibitemOpen
  \bibfield  {author} {\bibinfo {author} {\bibfnamefont {N.~A.}\ \bibnamefont {Rubin}}, \bibinfo {author} {\bibfnamefont {G.}~\bibnamefont {D'Aversa}}, \bibinfo {author} {\bibfnamefont {P.}~\bibnamefont {Chevalier}}, \bibinfo {author} {\bibfnamefont {Z.}~\bibnamefont {Shi}}, \bibinfo {author} {\bibfnamefont {W.~T.}\ \bibnamefont {Chen}},\ and\ \bibinfo {author} {\bibfnamefont {F.}~\bibnamefont {Capasso}},\ }\href {https://doi.org/10.1126/science.aax1839} {\bibfield  {journal} {\bibinfo  {journal} {Science}\ }\textbf {\bibinfo {volume} {365}},\ \bibinfo {pages} {eaax1839} (\bibinfo {year} {2019})}\BibitemShut {NoStop}%
\bibitem [{\citenamefont {Rubin}\ \emph {et~al.}(2021)\citenamefont {Rubin}, \citenamefont {Shi},\ and\ \citenamefont {Capasso}}]{Rubin2021a}%
  \BibitemOpen
  \bibfield  {author} {\bibinfo {author} {\bibfnamefont {N.~A.}\ \bibnamefont {Rubin}}, \bibinfo {author} {\bibfnamefont {Z.}~\bibnamefont {Shi}},\ and\ \bibinfo {author} {\bibfnamefont {F.}~\bibnamefont {Capasso}},\ }\href {https://doi.org/10.1364/AOP.439986} {\bibfield  {journal} {\bibinfo  {journal} {Adv. Opt. Photon.}\ }\textbf {\bibinfo {volume} {13}},\ \bibinfo {pages} {836} (\bibinfo {year} {2021})}\BibitemShut {NoStop}%
\bibitem [{\citenamefont {Rubin}\ \emph {et~al.}(2022)\citenamefont {Rubin}, \citenamefont {Chevalier}, \citenamefont {Juhl}, \citenamefont {Tamagnone}, \citenamefont {Chipman},\ and\ \citenamefont {Capasso}}]{Rubin2022}%
  \BibitemOpen
  \bibfield  {author} {\bibinfo {author} {\bibfnamefont {N.~A.}\ \bibnamefont {Rubin}}, \bibinfo {author} {\bibfnamefont {P.}~\bibnamefont {Chevalier}}, \bibinfo {author} {\bibfnamefont {M.}~\bibnamefont {Juhl}}, \bibinfo {author} {\bibfnamefont {M.}~\bibnamefont {Tamagnone}}, \bibinfo {author} {\bibfnamefont {R.}~\bibnamefont {Chipman}},\ and\ \bibinfo {author} {\bibfnamefont {F.}~\bibnamefont {Capasso}},\ }\href {https://doi.org/10.1364/oe.450941} {\bibfield  {journal} {\bibinfo  {journal} {Optics Express}\ }\textbf {\bibinfo {volume} {30}},\ \bibinfo {pages} {9389} (\bibinfo {year} {2022})}\BibitemShut {NoStop}%
\bibitem [{\citenamefont {Zaidi}\ \emph {et~al.}(2024)\citenamefont {Zaidi}, \citenamefont {Rubin}, \citenamefont {Meretska}, \citenamefont {Li}, \citenamefont {Dorrah}, \citenamefont {Park},\ and\ \citenamefont {Capasso}}]{Zaidi2024}%
  \BibitemOpen
  \bibfield  {author} {\bibinfo {author} {\bibfnamefont {A.}~\bibnamefont {Zaidi}}, \bibinfo {author} {\bibfnamefont {N.~A.}\ \bibnamefont {Rubin}}, \bibinfo {author} {\bibfnamefont {M.~L.}\ \bibnamefont {Meretska}}, \bibinfo {author} {\bibfnamefont {L.~W.}\ \bibnamefont {Li}}, \bibinfo {author} {\bibfnamefont {A.~H.}\ \bibnamefont {Dorrah}}, \bibinfo {author} {\bibfnamefont {J.-S.}\ \bibnamefont {Park}},\ and\ \bibinfo {author} {\bibfnamefont {F.}~\bibnamefont {Capasso}},\ }\href@noop {} {\bibfield  {journal} {\bibinfo  {journal} {Nature Photonics}\ }\textbf {\bibinfo {volume} {18}},\ \bibinfo {pages} {704} (\bibinfo {year} {2024})}\BibitemShut {NoStop}%
\bibitem [{\citenamefont {Yousef}\ \emph {et~al.}(2025)\citenamefont {Yousef}, \citenamefont {D’Alessandro}, \citenamefont {Yeh}, \citenamefont {Sinclair}, \citenamefont {Loncar},\ and\ \citenamefont {Capasso}}]{Yousef2025}%
  \BibitemOpen
  \bibfield  {author} {\bibinfo {author} {\bibfnamefont {K.~M.~A.}\ \bibnamefont {Yousef}}, \bibinfo {author} {\bibfnamefont {M.}~\bibnamefont {D’Alessandro}}, \bibinfo {author} {\bibfnamefont {M.}~\bibnamefont {Yeh}}, \bibinfo {author} {\bibfnamefont {N.}~\bibnamefont {Sinclair}}, \bibinfo {author} {\bibfnamefont {M.}~\bibnamefont {Loncar}},\ and\ \bibinfo {author} {\bibfnamefont {F.}~\bibnamefont {Capasso}},\ }\href {https://doi.org/10.1126/science.adw8404} {\bibfield  {journal} {\bibinfo  {journal} {Science}\ }\textbf {\bibinfo {volume} {389}},\ \bibinfo {pages} {416} (\bibinfo {year} {2025})},\ \Eprint {https://arxiv.org/abs/https://www.science.org/doi/pdf/10.1126/science.adw8404} {https://www.science.org/doi/pdf/10.1126/science.adw8404} \BibitemShut {NoStop}%
\bibitem [{\citenamefont {Li}\ \emph {et~al.}(2025)\citenamefont {Li}, \citenamefont {Oh}, \citenamefont {Miller}, \citenamefont {Capasso},\ and\ \citenamefont {Rubin}}]{Li2025}%
  \BibitemOpen
  \bibfield  {author} {\bibinfo {author} {\bibfnamefont {L.~W.}\ \bibnamefont {Li}}, \bibinfo {author} {\bibfnamefont {J.}~\bibnamefont {Oh}}, \bibinfo {author} {\bibfnamefont {H.}~\bibnamefont {Miller}}, \bibinfo {author} {\bibfnamefont {F.}~\bibnamefont {Capasso}},\ and\ \bibinfo {author} {\bibfnamefont {N.~A.}\ \bibnamefont {Rubin}},\ }\href@noop {} {\bibfield  {journal} {\bibinfo  {journal} {Optica}\ }\textbf {\bibinfo {volume} {12}},\ \bibinfo {pages} {799} (\bibinfo {year} {2025})}\BibitemShut {NoStop}%
\bibitem [{\citenamefont {Li}\ \emph {et~al.}(2026)\citenamefont {Li}, \citenamefont {Oakley}, \citenamefont {Schindhelm}, \citenamefont {Sellers}, \citenamefont {Casini},\ and\ \citenamefont {Rubin}}]{Li2026}%
  \BibitemOpen
  \bibfield  {author} {\bibinfo {author} {\bibfnamefont {L.~W.}\ \bibnamefont {Li}}, \bibinfo {author} {\bibfnamefont {P.~H.~H.}\ \bibnamefont {Oakley}}, \bibinfo {author} {\bibfnamefont {R.~N.}\ \bibnamefont {Schindhelm}}, \bibinfo {author} {\bibfnamefont {S.~G.}\ \bibnamefont {Sellers}}, \bibinfo {author} {\bibfnamefont {R.}~\bibnamefont {Casini}},\ and\ \bibinfo {author} {\bibfnamefont {N.~A.}\ \bibnamefont {Rubin}},\ }\href {https://doi.org/10.1126/sciadv.aee8035} {\bibfield  {journal} {\bibinfo  {journal} {Science Advances}\ }\textbf {\bibinfo {volume} {12}},\ \bibinfo {pages} {eaee8035} (\bibinfo {year} {2026})},\ \Eprint {https://arxiv.org/abs/https://www.science.org/doi/pdf/10.1126/sciadv.aee8035} {https://www.science.org/doi/pdf/10.1126/sciadv.aee8035} \BibitemShut {NoStop}%
\bibitem [{\citenamefont {Jones}(1941)}]{Jones1941}%
  \BibitemOpen
  \bibfield  {author} {\bibinfo {author} {\bibfnamefont {R.~C.}\ \bibnamefont {Jones}},\ }\href@noop {} {\bibfield  {journal} {\bibinfo  {journal} {Journal of the Optical Society of America}\ }\textbf {\bibinfo {volume} {31}},\ \bibinfo {pages} {488} (\bibinfo {year} {1941})}\BibitemShut {NoStop}%
\bibitem [{\citenamefont {Damask}(2005)}]{Damask2005}%
  \BibitemOpen
  \bibfield  {author} {\bibinfo {author} {\bibfnamefont {J.~N.}\ \bibnamefont {Damask}},\ }\href@noop {} {\emph {\bibinfo {title} {Polarization {{Optics}} in {{Telecommunications}}}}}\ (\bibinfo  {publisher} {Springer},\ \bibinfo {year} {2005})\BibitemShut {NoStop}%
\bibitem [{\citenamefont {Chipman}\ \emph {et~al.}(2019)\citenamefont {Chipman}, \citenamefont {Lam},\ and\ \citenamefont {Young}}]{Chipman2019}%
  \BibitemOpen
  \bibfield  {author} {\bibinfo {author} {\bibfnamefont {R.~A.}\ \bibnamefont {Chipman}}, \bibinfo {author} {\bibfnamefont {W.-S.~T.}\ \bibnamefont {Lam}},\ and\ \bibinfo {author} {\bibfnamefont {G.}~\bibnamefont {Young}},\ }\href@noop {} {\emph {\bibinfo {title} {Polarized {{Light}} and {{Optical Systems}}}}}\ (\bibinfo  {publisher} {CRC Press},\ \bibinfo {year} {2019})\BibitemShut {NoStop}%
\bibitem [{\citenamefont {Molesky}\ \emph {et~al.}(2018)\citenamefont {Molesky}, \citenamefont {Lin}, \citenamefont {Piggott}, \citenamefont {Jin}, \citenamefont {Vuckovi{\'c}},\ and\ \citenamefont {Rodriguez}}]{Molesky2018}%
  \BibitemOpen
  \bibfield  {author} {\bibinfo {author} {\bibfnamefont {S.}~\bibnamefont {Molesky}}, \bibinfo {author} {\bibfnamefont {Z.}~\bibnamefont {Lin}}, \bibinfo {author} {\bibfnamefont {A.~Y.}\ \bibnamefont {Piggott}}, \bibinfo {author} {\bibfnamefont {W.}~\bibnamefont {Jin}}, \bibinfo {author} {\bibfnamefont {J.}~\bibnamefont {Vuckovi{\'c}}},\ and\ \bibinfo {author} {\bibfnamefont {A.~W.}\ \bibnamefont {Rodriguez}},\ }\href@noop {} {\bibfield  {journal} {\bibinfo  {journal} {Nature photonics}\ }\textbf {\bibinfo {volume} {12}},\ \bibinfo {pages} {659} (\bibinfo {year} {2018})}\BibitemShut {NoStop}%
\bibitem [{\citenamefont {Lalau-Keraly}\ \emph {et~al.}(2013)\citenamefont {Lalau-Keraly}, \citenamefont {Bhargava}, \citenamefont {Miller},\ and\ \citenamefont {Yablonovitch}}]{Lalau-Keraly2013}%
  \BibitemOpen
  \bibfield  {author} {\bibinfo {author} {\bibfnamefont {C.~M.}\ \bibnamefont {Lalau-Keraly}}, \bibinfo {author} {\bibfnamefont {S.}~\bibnamefont {Bhargava}}, \bibinfo {author} {\bibfnamefont {O.~D.}\ \bibnamefont {Miller}},\ and\ \bibinfo {author} {\bibfnamefont {E.}~\bibnamefont {Yablonovitch}},\ }\href {https://doi.org/10.1364/OE.21.021693} {\bibfield  {journal} {\bibinfo  {journal} {Opt. Express}\ }\textbf {\bibinfo {volume} {21}},\ \bibinfo {pages} {21693} (\bibinfo {year} {2013})}\BibitemShut {NoStop}%
\bibitem [{\citenamefont {Miller}(2012)}]{Miller2012}%
  \BibitemOpen
  \bibfield  {author} {\bibinfo {author} {\bibfnamefont {O.~D.}\ \bibnamefont {Miller}},\ }{\selectlanguage {English}\emph {\bibinfo {title} {Photonic Design: From Fundamental Solar Cell Physics to Computational Inverse Design}}},\ \href {https://www.proquest.com/dissertations-theses/photonic-design-fundamental-solar-cell-physics/docview/1081752143/se-2} {Ph.D. thesis},\ \bibinfo  {school} {University of California Berkeley} (\bibinfo {year} {2012})\BibitemShut {NoStop}%
\bibitem [{\citenamefont {Sell}\ \emph {et~al.}(2017)\citenamefont {Sell}, \citenamefont {Yang}, \citenamefont {Doshay}, \citenamefont {Yang},\ and\ \citenamefont {Fan}}]{Sell2017}%
  \BibitemOpen
  \bibfield  {author} {\bibinfo {author} {\bibfnamefont {D.}~\bibnamefont {Sell}}, \bibinfo {author} {\bibfnamefont {J.}~\bibnamefont {Yang}}, \bibinfo {author} {\bibfnamefont {S.}~\bibnamefont {Doshay}}, \bibinfo {author} {\bibfnamefont {R.}~\bibnamefont {Yang}},\ and\ \bibinfo {author} {\bibfnamefont {J.~A.}\ \bibnamefont {Fan}},\ }\href {https://doi.org/10.1021/acs.nanolett.7b01082} {\bibfield  {journal} {\bibinfo  {journal} {Nano Letters}\ }\textbf {\bibinfo {volume} {17}},\ \bibinfo {pages} {3752} (\bibinfo {year} {2017})}\BibitemShut {NoStop}%
\bibitem [{\citenamefont {Liu}\ \emph {et~al.}(2025)\citenamefont {Liu}, \citenamefont {Wu}, \citenamefont {Lin}, \citenamefont {Su}, \citenamefont {Zhou},\ and\ \citenamefont {Huang}}]{Liu2025}%
  \BibitemOpen
  \bibfield  {author} {\bibinfo {author} {\bibfnamefont {Y.-T.}\ \bibnamefont {Liu}}, \bibinfo {author} {\bibfnamefont {Y.-C.}\ \bibnamefont {Wu}}, \bibinfo {author} {\bibfnamefont {C.-Y.}\ \bibnamefont {Lin}}, \bibinfo {author} {\bibfnamefont {H.-T.}\ \bibnamefont {Su}}, \bibinfo {author} {\bibfnamefont {Y.-Q.}\ \bibnamefont {Zhou}},\ and\ \bibinfo {author} {\bibfnamefont {Y.-W.}\ \bibnamefont {Huang}},\ }\href@noop {} {\bibfield  {journal} {\bibinfo  {journal} {Advanced Optical Materials}\ }\textbf {\bibinfo {volume} {13}},\ \bibinfo {pages} {2500872} (\bibinfo {year} {2025})}\BibitemShut {NoStop}%
\bibitem [{\citenamefont {Devlin}\ \emph {et~al.}(2016)\citenamefont {Devlin}, \citenamefont {Khorasaninejad}, \citenamefont {Chen}, \citenamefont {Oh},\ and\ \citenamefont {Capasso}}]{Devlin2016}%
  \BibitemOpen
  \bibfield  {author} {\bibinfo {author} {\bibfnamefont {R.~C.}\ \bibnamefont {Devlin}}, \bibinfo {author} {\bibfnamefont {M.}~\bibnamefont {Khorasaninejad}}, \bibinfo {author} {\bibfnamefont {W.-T.}\ \bibnamefont {Chen}}, \bibinfo {author} {\bibfnamefont {J.}~\bibnamefont {Oh}},\ and\ \bibinfo {author} {\bibfnamefont {F.}~\bibnamefont {Capasso}},\ }\href {https://doi.org/10.1073/pnas.1611740113} {\bibfield  {journal} {\bibinfo  {journal} {Proceedings of the National Academy of Sciences}\ }\textbf {\bibinfo {volume} {113}},\ \bibinfo {pages} {10473} (\bibinfo {year} {2016})}\BibitemShut {NoStop}%
\end{thebibliography}
\end{document}